\documentclass[11pt]{article}
\usepackage{amssymb,amsmath,cite}
\catcode`\@=11

\long\def\@makefntext#1{\parindent 0cm\noindent
\hbox to 1em{\hss$^{\@thefnmark}$}#1}

\newcommand{\captionfonts}{\small}
\makeatletter  
\long\def\@makecaption#1#2{%
  \vskip\abovecaptionskip
  \sbox\@tempboxa{{\captionfonts #1: #2}}%
  \ifdim \wd\@tempboxa >\hsize
    {\captionfonts #1: #2\par}
  \else
    \hbox to\hsize{\hfil\box\@tempboxa\hfil}%
  \fi
  \vskip\belowcaptionskip}
\makeatother   

\newcommand{\beq}{\begin{equation}}
\newcommand{\eeq}{\end{equation}}

\begin{document}
\begin{titlepage}
\vspace{.5in}
\begin{flushright}
UCD-05-09\\
gr-qc/0508072\\
July 2005\\
\end{flushright}
\vspace{.5in}
\begin{center}
{\Large\bf
 A Note on Real Tunneling Geometries}\\
\vspace{.4in}
{S.~C{\sc arlip}\footnote{\it email: carlip@physics.ucdavis.edu}\\
       {\small\it Department of Physics}\\
       {\small\it University of California}\\
       {\small\it Davis, CA 95616}\\{\small\it USA}}
\end{center}

\vspace{.5in}
\begin{center}
{\large\bf Abstract}
\end{center}
\begin{center}
\begin{minipage}{4.5in}
{\small
In the Hartle-Hawking ``no boundary'' approach to quantum cosmology, a 
real tunneling geometry is a configuration that represents a transition
from a compact Riemannian spacetime to a Lorentzian universe.  I complete
an earlier proof that in three spacetime dimensions, such a transition 
is ``probable,'' in the sense that the required Riemannian geometry yields 
a genuine maximum of the semiclassical wave function.
}
\end{minipage}
\end{center}
\end{titlepage}
\addtocounter{footnote}{-1}
\setlength{\textfloatsep}{1.5ex}
\begin{figure} 
\begin{picture}(100,105)(-150,12)
\qbezier(21,20)(-5,49)(16,90)
\qbezier(95,20)(120,49)(102,92)
\qbezier(21,20)(58,-11)(95,20)
\qbezier(16,90)(20,108)(12,125)
\qbezier(102,92)(98,110)(106,125)
\qbezier(12,125)(60,105)(106,125)
\qbezier(12,125)(60,142)(106,125)
\put(54,121){$\Sigma$}
\put(50,40){$M$}
\qbezier[40](16,89)(60,66)(103,89)
\qbezier[40](16,89)(60,110)(103,89)
\put(53,85){$\Sigma_0$}
\end{picture}
\caption{\small A manifold $M$ with a single boundary $\Sigma$ describes the
birth of a universe in the Hartle--Hawking approach to quantum cosmology.
In a real tunneling geometry, the signature of the metric changes from
Riemannian to Lorentzian across an intermediate hypersurface $\Sigma_0$.
\label{figa}}
\end{figure}
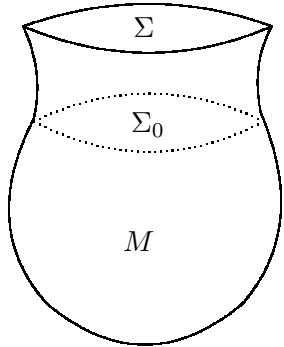
 
In the Hartle-Hawking approach to quantum cosmology \cite{Hawking,HartHawk},
the Universe is described by a Euclidean path integral on an $n$-dimensional 
manifold $M$ with a single boundary component $\Sigma$ representing the 
``present'' (see figure \ref{figa}).  The path integral depends on the 
induced metric $h_{ij}$ on $\Sigma$ and the boundary values of any matter 
fields $\varphi$ on $\Sigma$, thus yielding a ``wave function of the Universe'' 
$\Psi[h_{ij},\varphi|_\Sigma]$.  Whether one should also sum over topologies 
of $M$ is an open question; such a sum can qualitatively change the locations 
of the peaks of the wave function \cite{Carlipa,ACRST}, but does not affect 
the main thrust of this paper.

In the absence of a full-fledged quantum theory of gravity, of course, such 
a path integral is not very well defined.  The hope is that minisuperspace
models and semiclassical saddle point approximations might still give useful
information about, for example, inflation \cite{Hawkingb}.  If $\Sigma$ is
to be spacelike, one cannot ordinarily find a saddle point with a globally
Lorentzian metric---such topology-changing geometries are forbidden by fairly 
mild energy conditions \cite{Tipler}.  It may be that the dominant saddle points
are complex \cite{Halliwell}, with consequent ambiguities in the integration
contour.  If we restrict ourselves to real metrics, though, we are naturally 
led to ``real tunneling geometries'' \cite{GibHart}, geometries in which an 
initial Riemannian metric is joined to a Lorentzian metric along a hypersurface 
$\Sigma_0$, as shown in figure \ref{figa}.

For the resulting geometry to be smooth, with no ``boundary layer'' stress-%
energy tensor at $\Sigma_0$, the induced metrics $h_{ij}$ must match across 
$\Sigma_0$ and the extrinsic curvature $K_{ij}[\Sigma_0]$ of the signature-%
changing hypersurface must vanish.  A classical solution of the field equations 
with such a geometry may or may not exist, depending on the topology of $M$ 
and the sign of the cosmological constant.  Some of the known restrictions 
on the topology of $M$ are described in \cite{GibHart} and \cite{Gibbons}.  

For the special case of a three-manifold $M$ with a negative cosmological 
constant, the question of which topologies admit real tunneling geometries 
is almost completely solved.  Any three-dimensional Einstein metric with 
$\Lambda<0$ is hyperbolic (that is, has constant negative curvature).  
Thurston has shown that a compact three-manifold with a nontrivial boundary 
admits a hyperbolic metric if and only if it is prime, homotopically atoroidal, 
and not homeomorphic to a certain twisted product of a two-torus and an interval 
\cite{Thurston}.  If in addition $M$ is acylindrical, it admits a hyperbolic 
metric for which the extrinsic curvature of the boundary vanishes, and thus 
allows a real tunneling geometry.

Real tunneling geometries lead to an elegant classical picture of a universe 
born from ``nothing.''  Quantum mechanically, though, the picture is less clear.  
The induced metric and extrinsic curvature are conjugate variables, and in a 
quantum theory, they should not be specified simultaneously.  In the saddle 
point approximation, in particular, the requirement that $K_{ij}[\Sigma_0]=0$ 
determines the boundary metric $h_{ij}$ nearly uniquely.  Hence real tunneling 
geometries do not determine a wave function $\Psi[h_{ij}]$, but merely a 
contribution to the wave function at a few particular values of $h_{ij}$.

One alternative, proposed in \cite{Carlip}, is to forget for a moment about 
the requirement that $K_{ij}[\Sigma_0]=0$, and consider the wave function 
$\Psi[h_{ij}]$ as a functional of the spatial metric on $\Sigma_0$.  We can 
then ask whether a real tunneling geometry is \emph{probable}---that is, whether 
configurations with $K_{ij}=0$ occur at or near the peaks of the wave function.  
This is still a bit tricky, since the spatial metric contains information about 
time as well as spatial geometry \cite{Wheeler}; we should really ask
whether a transition to Lorentzian signature is probable at a fixed time.
Following York \cite{York}, we can use the mean curvature $K=h^{ij}K_{ij}$ as 
an ``extrinsic time'' coordinate.  The wave function will then be a functional 
of $K$ and the conformal metric ${\tilde h}_{ij}$, that is, the spatial metric 
modulo Weyl transformations \cite{Isenberg,Fischer}.  

The question is then whether configurations with $K_{ij}=0$ are probable at 
time $K=0$.  To answer this in the semiclassical approximation, we must first 
find a form of the Einstein-Hilbert action that fixes $K$ and ${\tilde h}_{ij}$ 
at the boundary.  (For simplicity, I will omit matter terms.)  Recall first 
\cite{Hawking,Carlip} that the Euclidean action 
\beq
I_E[g] = -\frac{1}{16\pi G}\int_M d^n\!x\sqrt{g}(R-2\Lambda)
         -\frac{1}{8\pi G}\int_{\Sigma_0} d^{n-1}\!x\sqrt{h} K
\label{a1}
\eeq
is appropriate for a fixed boundary metric $h_{ij}$:
\beq
\delta I_E = ({\mathit equations\, of\, motion})
  -\frac{1}{16\pi G}\int_{\Sigma_0}d^{n-1}\!x \sqrt{h}(K_{ij}-h_{ij}K)\delta h^{ij}
 \label{a2}  
\eeq
so no boundary contributions appear in the variation when $\delta h^{ij}$ 
vanishes.  Adding a term
$$\frac{1}{8\pi G}\frac{n-2}{n-1}\int_{\Sigma_0}d^{n-1}\!x \sqrt{h}K$$
to (\ref{a1}), we obtain a new ``York time'' action
\beq
I_Y[g] = -\frac{1}{16\pi G}\int_M d^n\!x\sqrt{g}(R[g]-2\Lambda)
         -\frac{1}{8\pi G(n-1)}\int_{\Sigma_0} d^{n-1}\!x\sqrt{h} K ,
\label{a3}
\eeq
whose variation is
\begin{align}
\delta I_Y &= ({\mathit equations\, of\, motion})\label{a4}\\ 
  &-\frac{1}{16\pi G}\int_{\Sigma_0}d^{n-1}\!x 
  \sqrt{h}(K_{ij}-\frac{1}{n-1} h_{ij}K)\delta h^{ij} 
  + \frac{1}{8\pi G}\frac{n-2}{n-1}\int_{\Sigma_0}d^{n-1}\!x\sqrt{h}\delta K .
\nonumber 
\end{align}
For a fixed conformal geometry on $\Sigma_0$, the only allowed metric variations  
are of the form $\delta h^{ij} = \delta\phi\, h^{ij}$, so the first boundary term
in (\ref{a4}) vanishes; for $K$ fixed, the second term vanishes as well.

Saddle point contributions to the path integral come from extrema of (\ref{a3}),
that is, classical solutions of the Einstein field equations with prescribed mean 
curvature and conformal geometry at $\Sigma_0$.  For a solution ${\bar g}_{ab}$
with $K[\Sigma_0]=0$, the action (\ref{a3}) is
\beq
{\bar I}_Y[{\bar g}] = -\frac{\Lambda}{4\pi G(n-2)}{\mathop{Vol}}_{\bar g}(M) ,
\label{a5}
\eeq
and the saddle point contribution to the path integral is
\beq
\Psi[{\bar h}_{ij},K=0] \sim \Delta_{\bar g}
  \exp\left\{\frac{\Lambda}{4\pi G(n-2)}{\mathop{Vol}}_{\bar g}(M)\right\} ,
\label{a6}
\eeq
where the Van Vleck-Morette determinant $\Delta_{\bar g}$ is a combination of 
determinants coming from quadratic terms in the action and from gauge-fixing.

It is immediately evident from (\ref{a4}) that the extrema of the classical
action ${\bar I}_Y$ occur at $K_{ij}[\Sigma_0]=0$: when one varies the boundary 
metric\footnote{Note that I am considering only variations among classical 
solutions.  Variations of the conformal factor off the space of solutions can 
make the action arbitrarily negative \cite{Hawking}.  There is evidence that 
such variations are unimportant in the full path integral \cite{Mottola,Loll,Lollb}, 
but the question is not yet settled.}  while keeping $K[\Sigma_0]=0$, only the 
second term on the right-hand side of (\ref{a4}) contributes to $\delta{\bar I}_Y$. 
The question of whether these extrema are actually minima was first addressed 
in \cite{Carlip}.  In more than three spacetime dimensions, the answer is not 
known: the second variation of the action includes a term proportional to the 
Weyl tensor, whose contribution I do not know how to control.  

In three spacetime dimensions, though, the Weyl tensor term is absent, and the 
problem is more tractable.  It was shown in \cite{Carlip} that for $\Lambda>0$, 
extrema with vanishing extrinsic curvature at $\Sigma_0$ are local minima of 
${\bar I}_Y$, and thus---assuming that the Van Vleck-Morette contribution is 
small---local maxima of the wave function (\ref{a6}).  For this case, then, 
real tunneling geometries are ``probable.''

The main aim of this paper is to extend this analysis to geometries for 
which $\Lambda<0$, a case left unresolved in \cite{Carlip}.  This extension
is made possible by a new result of Agol, Storm, and Thurston \cite{ADST}, 
who prove---assuming the correctness of Perelman's recent work on on the 
geometrization theorem \cite{Perelman}---the following:
\begin{center}
\begin{minipage}{5.7in} 
Let $(M,g)$ be a compact hyperbolic three-manifold with a minimal surface 
[i.e., $K=0$] boundary.  If $M$ is acylindrical, it admits a hyperbolic 
metric $\nu$ with a totally geodesic [i.e., $K_{ij}=0$] boundary.  Then 
$\mathop{Vol}(M,g)\ge\mathop{Vol}(M,\nu)$.
\end{minipage}
\end{center}
In other words, given a topological restriction on $M$ (acylindricity) that 
guarantees the existence of a hyperbolic metric for which $K_{ij}[\Sigma_0]=0$,
the geometry for which $K_{ij}[\Sigma_0]=0$ has minimal volume among all 
hyperbolic geometries for which $K[\Sigma_0]=0$.  But all 
vacuum solutions of the Einstein field equations in three dimensions with 
$\Lambda<0$ are hyperbolic, and when $\Lambda<0$, the smallest volumes give 
the largest contributions to the wave function (\ref{a6}).  We thus conclude%
---again assuming that the Van Vleck-Morette contribution is small---that real
tunneling geometries are ``probable'' for $\Lambda<0$ as well.

Thus far, I have assumed that the determinant $\Delta_{\bar g}$ in (\ref{a6}) 
is unimportant in determining the peaks of the wave function.  In three 
spacetime dimensions, this factor is essentially the Ray-Singer torsion 
\cite{Witten}, which can be computed in certain cases (see, for example, 
the appendix of \cite{Carlipa}).  For a manifold with boundary, the dependence 
on the extrinsic curvature---or, roughly equivalently, on the boundary spin 
connection in the Chern-Simons formalism---is analyzed carefully in \cite{CarCos}.  
In principle, it should be possible to use this result to test the assumption 
that $\Delta_{\bar g}$ can be neglected here.

To extend these results to more that three dimensions requires control of the
Weyl curvature term discussed in \cite{Carlip}.  But the success in three
dimensions at least makes the Hartle-Hawking description of quantum tunneling 
from Riemannian to Lorentzian signature more plausible.

\vspace{1.5ex}
\begin{flushleft}
\large\bf Acknowledgments
\end{flushleft}

\noindent I would like to thank Peter Storm for pointing out and helping
explain his and his collaborators' recent results.  This work was supported 
in part by Department of Energy grant DE-FG02-91ER40674.

\end{document}